\documentclass[11pt,a4paper]{article}
\pdfoutput=1
\usepackage{jcappub}
\usepackage{graphicx}
\usepackage{wasysym}
\usepackage{subfig}
\usepackage{multirow}
\usepackage{array}

\usepackage{tabularx}
\usepackage[table]{xcolor}

  \title{Constraints on the Tensor-to-Scalar ratio for  non-power-law models}

\author[a,b]{J.~Alberto V\'azquez}
\author[a,b]{M.~Bridges}
 \author[c,d]{Yin-Zhe Ma}
\author[b]{M.P.~Hobson}

\affiliation[a]{Kavli Institute for Cosmology, Madingley Road,
Cambridge CB3 0HA, UK.} \affiliation[b]{Astrophysics Group,
Cavendish Laboratory, JJ Thomson Avenue, Cambridge, CB3 0HE, UK.}
\affiliation[c]{Department of Physics and Astronomy, University of
British Columbia, Vancouver, V6T 1Z1, BC Canada.}
\affiliation[d]{Canadian Institute for Theoretical Astrophysics,
Toronto, M5S 3H8, Ontario, Canada.}

\emailAdd{jv292@cam.ac.uk}
\arxivnumber{1303.4014}

\abstract{
Recent cosmological  observations hint at a
deviation from the simple power-law form of the primordial spectrum of curvature perturbations.
In this paper we  show that in the presence of a tensor component,
a  turn-over in the initial spectrum is preferred by current observations,
and hence non-power-law models ought to be considered.
For instance, for  a power-law parameterisation with both a tensor component 
and running parameter,
current data show a preference for a negative running at more than $2.5\sigma$ C.L.
As a consequence of this deviation from a power-law, constraints on the
tensor-to-scalar ratio $r$ are slightly broader.
We also present constraints on the inflationary parameters
for a model-independent reconstruction and the Lasenby \& Doran (LD) model.
In particular, the constraints on the tensor-to-scalar ratio from the LD model are:
  $r_{\rm LD}=0.11\pm{0.024}$.
In addition to current data, we show expected constraints  from 
Planck-like and CMB-Pol sensitivity experiments by using
Markov-Chain-Monte-Carlo sampling chains. For all the models, we
have included the Bayesian Evidence to perform a model selection
analysis.
 The Bayes factor, using  current observations, shows a strong 
 preference for the LD model over the standard power-law parameterisation,
and  provides an insight into the accuracy of differentiating models through future surveys.}

\keywords{Primordial Power Spectrum, Cosmological Parameters from CMBR,
  Inflation, Bayesian Analysis}
\begin{document}
  \maketitle
   \flushbottom

\section{Introduction}

Inflationary models have the merit that they not only explain the
 homogeneity of the universe on large-scales, but also provide a theory
to explain the origin of perturbations as observed in the Cosmic
Microwave Background (CMB). During the inflationary period,
quantum fluctuations of the field were driven to scales much
larger than the Hubble horizon and eventually turned into  
density perturbations (scalar) observed in the CMB and the large-scale distribution of galaxies,
together with a gravitational wave (tensor) contribution.
%
The vector contributions, however, are  expected to be negligible 
since these modes decayed very rapidly once they entered the Hubble horizon.
The scalar  and tensor contributions are summarised by the primordial power 
spectrum $\mathcal{P_R}(k)$ and $\mathcal{P_T}(k)$, respectively.
The scalar spectrum for a single-inflaton field $\phi$, in the slow-roll approximation,
 is  given by  \cite{Liddle92}:

\begin{equation} \label{spec1}
  \mathcal{P_R}(k) =\left[ \left(\frac{H}{\dot \phi}\right)^2 \left(\frac{H}{2\pi}\right)^2 \right]_{k=aH},
\end{equation}

\noindent where the expression is evaluated at the  horizon exit
$k=aH$. Since the initial spectrum is an unknown function, one
needs to carry out a  full numerical calculation from the onset of
the inflationary phase,  or to assume a particular functional form
of it. The simplest proposal is to parameterise the shape of
$\mathcal{P_R}(k)$ by a power-law.
Although the power-law assumption has provided reasonable agreement
with cosmological observations, some recent analyses
have shown that if a running of the scalar spectral-index is taken into account,
there exists a preference for a negative running-value at $1.8\sigma$ C.L. 
from  WMAP7+ACT (SPT) measurements \cite{ACT,SPT} and at 2.2$\sigma$ C.L. 
with WMAP7$+$QUaD~\cite{QUAD}.
It has also been shown that the existence of a turn-over in
$\mathcal{P_R}(k)$, by
using model-independent techniques, is preferred
\cite{Parkinson10, Guo11, Guo11b, Vazquez12}. The presence of this turn-over
plays an important role  in explaining current cosmological
observations and  cannot therefore be ignored when constraining the 
inflationary parameters. In the slow-roll approximation, the shape of 
the spectrum of tensor perturbations is 
\begin{equation}\label{spec2}
   \mathcal{P_T}(k) =\left[  \frac{16}{\pi}H^2 \right]_{k=aH},
\end{equation}

\noindent
which depends on the form of the scalar spectrum, and vice-versa, via the potential of 
the single scalar-field.
To place constraints on the amplitude of tensor contributions, it
is customary to define the tensor-to-scalar ratio as

\begin{equation}\label{eq:rat}
r(k)\equiv \frac{\mathcal{P_T}(k)}{\mathcal{P_R}(k)}=64\pi\left( \frac{\dot \phi^2}{H^2} \right)_{k=aH}.
\end{equation}

\noindent 
The dependence of the scalar spectrum on the tensor
spectrum is evident in the Lasenby \& Doran  model \cite{Lasenby05}, where
both spectra depend upon the same best-fit parameters.
In a previous paper we found  that  standard $\Lambda$CDM models
with a turn-over in the scalar spectrum are preferred over a
simple power-law parameterisation \cite{Vazquez12}. In this work, by assuming
a power-law parameterisation of the tensor spectrum, we show that the bending of
the scalar spectrum is  enhanced due to the presence of a tensor
component. To avoid misleading  results  due to the particular
choice  of parameterisation, the shape of the scalar
spectrum is described by employing a model-independent reconstruction. We then
show that current constraints on the tensor-to-scalar ratio
(\ref{eq:rat}) are broadened for non-power law $\mathcal{P_R}(k)$ models. We also
discuss the constraints on $r$ for a massive scalar-field in  the Lasenby \&  Doran model. 
Finally, by considering future
experiments we present their expected constraints on the
inflationary parameters. For all the models,  the
Bayes factor is computed in order to perform a model comparison.
\\

The paper is organised as follows: in the next Section we list the data sets
and the cosmological parameters considered. In Section~\ref{sec:spec}
we study different models suggested to describe the form of the scalar spectrum.
Then, we show the resulting parameter constraints on the 
tensor-to-scalar ratio
and the preferred form of the power spectrum using current cosmological observations.
In the same section we provide future constraints on $r$ expected by
Planck-like and CMB-Pol experiments.
Performance assumptions for Planck and CMB-Pol are taken from 
\cite{Planck} and \cite{CMBPol}.
We present the model selection analysis in Section~\ref{sec:mod_sel}, and  
our conclusions in Section~\ref{sec:results}.

\section{Theoretical Framework}
\label{sec:theor}

Even though the primary parameters in the standard $\Lambda$CDM
model have already been  tightly constrained and have little
impact on the $B$-mode spectrum, it is worthwhile to perform
a full parameter-space exploration to determine the
tensor-to-scalar ratio constraints in each model. We assume purely
Gaussian adiabatic scalar and  tensor contributions in a flat
$\Lambda$CDM model~\footnote{Except for the LD model, which is
based on a marginally closed universe $\Omega_k <0$.} specified by the
standard parameters: the physical baryon $\Omega_{\rm b} h^2$ and
cold dark matter density $\Omega_{\rm c} h^2$ relative to the
critical density ($h$ is the  dimensionless Hubble parameter such
that $H_0=100h$ kms$^{-1}$Mpc$^{-1}$), $\theta$ is $100 \times$
the ratio of the sound horizon to angular diameter distance at
last scattering surface, $\tau$ denotes the optical depth at
reionisation. We consider the tensor-to-scalar  ratio for each
model $i$ as $r_i=\mathcal{P_T}_{(i)}(k)/\mathcal{P_R}_{(i)}(k)$;
hereafter we set $r_i=r_i(k_0)$ at a
scale of $k_0= 0.015 {\rm Mpc^{-1}}$. A study of the appropriate
scale to use is  given  by \cite{Marina07}.
Aside from the Sunyaev-Zel'dovich (SZ) amplitude $A_{SZ}$ used by WMAP analyses,
the ACT likelihood incorporates two additional nuisance parameters:
the total Poisson power $A_p$ and the amplitude of the clustered power $A_c$.
 The parameters describing the primordial  spectra for each model are listed in the next
section, together with the flat priors imposed in our Bayesian analysis.
\\

Throughout the analysis,
the theoretical temperature and polarisation $C_\ell$'s spectra are generated
with a modified version of the CAMB code \cite{CAMB}, and the parameter estimation is
performed using the CosmoMC  program \cite{Cosmo}.
The calculation of the Bayesian evidence $\mathcal{Z}$, to perform the model selection,
requires a multidimensional integration over the likelihood and prior.
To do this, we make use of the {\sc MultiNest} algorithm \cite{Multi1, Multi2}.
The Bayes factor $\mathcal{B}_{ij}$, or equivalently the difference in
log evidences $\ln \mathcal{Z}_i - \ln \mathcal{Z}_j$, provides a
measure of how well model $i$ fits the data compared to model $j$.
In order to make a qualitative  model  comparison, we consider the
Jeffreys guideline: if $\mathcal{B}_{ij}< 1$ model $i$
should not be favoured over model $j$, $1<\mathcal{B}_{ij}<2.5$
constitutes significant evidence, $2.5<\mathcal{B}_{ij}<5$ is strong evidence, while
$\mathcal{B}_{ij}>5$ would be considered decisive \cite{Trotta08,Vazquez11}.

\subsection*{Current Observations}

\begin{figure}
\begin{center}
\includegraphics[trim = 1mm 1mm 2mm 1mm, clip, width=10 cm, height=5cm]{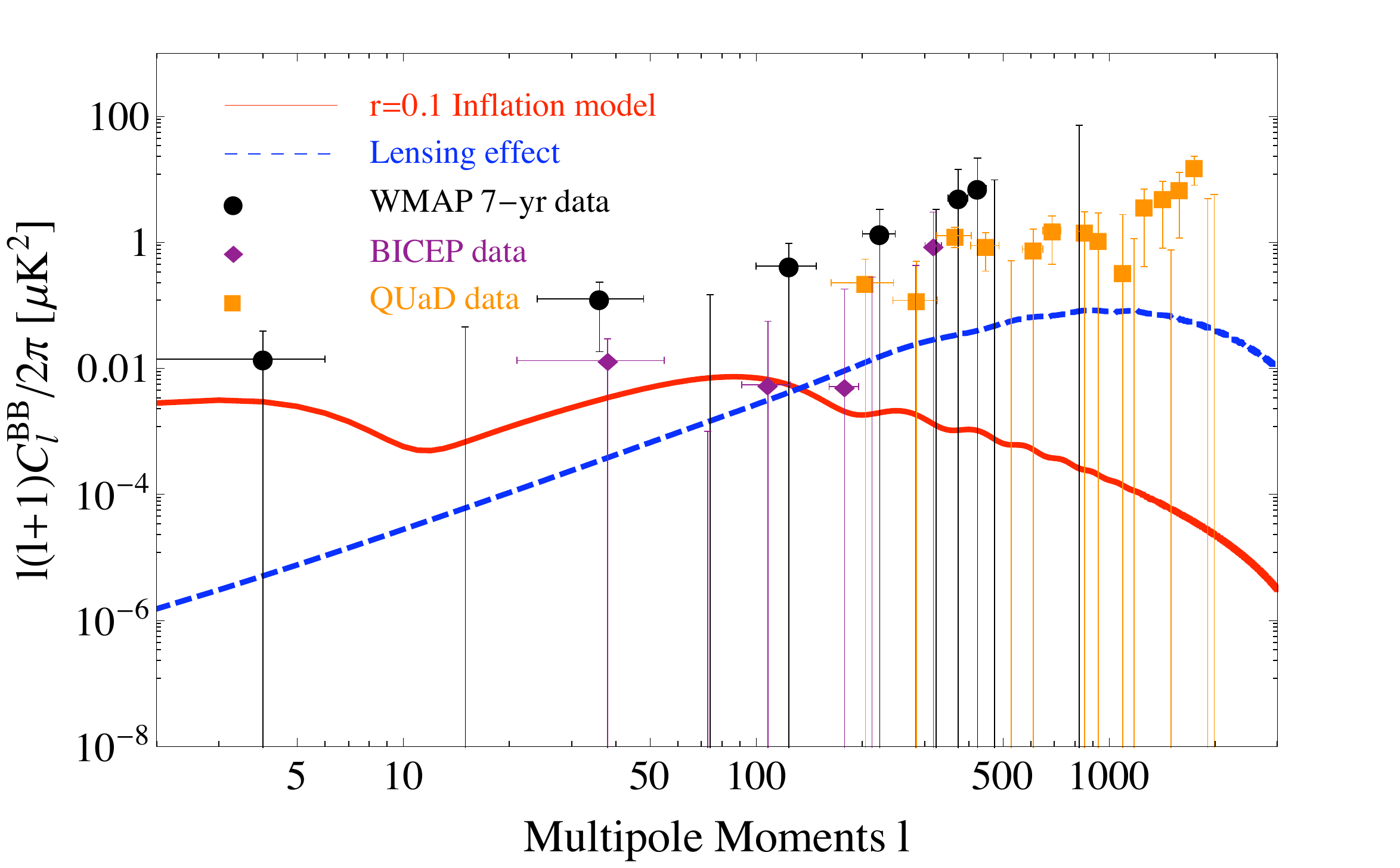}
 \caption{Comparison of theoretical prediction of $r=0.1$ inflation model, and the WMAP, 
BICEP and QUaD data for the $B$-mode power spectrum.}
 \label{fig:B_Pol}
\end{center}
\end{figure}

To compute posterior probabilities for each model
we use temperature and polarisation measurements  from the
 Wilkinson Microwave Anisotropy Probe 7-year (WMAP7; \cite{WMAP})
and the Atacama Cosmology Telescope (ACT; \cite{ACT}) data.
To improve polarisation constraints, we  include observations from
 QuaD \cite{QUAD}, whose primary aim is high resolution measurements ($154\le \ell \le 2026$)
of the $E$-mode signal, and BICEP data \cite{BICEP} which
probes intermediate scales $(21 \le \ell \le 335)$.
Figure \ref{fig:B_Pol} shows the $B$-mode spectrum predicted  from a power-law parameterisation,
with $r=0.1$, along with $1\sigma$ constraints obtained by using current observations.
In addition to CMB data, and to strengthen the  constraining power,  we incorporate 
distance measurements from the
Supernova Cosmology Project Union 2  compilation (SCP; \cite{SCP}) and Large Scale Structure 
data from the Sloan Digital Sky Survey (SDSS) Data Release 7 (DR7) Luminous Red
Galaxy (LRG) power spectrum \cite{LRG}. We also also consider baryon density information from
Big Bang Nucleosyntesis (BBN, \cite{BBN})  and impose a Gaussian prior
on the Hubble parameter today $H_0$  from  measurements of the  Hubble Space 
Telescope (HST; \cite{HST}) key project.

\subsection*{Future surveys}

In order to forecast expected constraints of future CMB observations, one can
make use of the Fisher information matrix under the assumption that each parameter is
Gaussian-distributed. However to obtain constraints more closely related to 
what is expected by future experiments, we perform a  Monte Carlo analysis
sampling over all variables involved in the description of the CMB spectrum.
We simulate future experiments by generating  mock data of the 
$\hat{C_\ell}^{XY}$'s from a $\chi^2_{2\ell+1}$  distribution with variances \cite{Powell11}:

   \begin{eqnarray}
 (\Delta \hat{C}_\ell^{XX})^2 &=& \frac{2}{(2\ell+1)f_{sky}}\left( C_\ell^{XX}+ N_\ell^{XX}\right)^2, \\
 (\Delta \hat{C}_\ell^{TE})^2 &=& \frac{2}{(2\ell+1)f_{sky}}\left[ \left( C_\ell^{TE} \right)^2+ \left(C_\ell^{TT}+ N_\ell^{TT}\right)
    \left(C_\ell^{EE}+N_\ell^{EE} \right) \right],
   \end{eqnarray}

   \noindent
   where $X=T,E$ and $B$ label the temperature and polarisations;
   $f_{sky}$ is the fraction of the observed sky. The $C_\ell^{XY}$'s
   represent the theoretical spectra and $N_\ell^{XY}$ the instrumental noise spectra 
   for each experiment. In experiments with multiple frequency channels $c$, 
   the noise spectrum is approximated \cite{Bowden04} by

  \begin{equation}
   N_\ell^X= \left(\sum_c \frac{1}{N_{\ell,c}^X}  \right)^{-1},
   \end{equation}

   \noindent
   where the noise spectrum of an individual frequency channel, assuming a Gaussian beam, is

   \begin{equation}
    N_{\ell,c}^X = (\sigma_{\rm pix}\, \theta_{\rm fwhm})^2 \exp \left[\ell(\ell+1)\frac{\theta^2_{\rm fwhm}}{8\ln2}\right]\delta_{XY}.
   \end{equation}

   \noindent
    The pixel noise from temperature and polarisation maps are considered as uncorrelated.
    The noise per pixel $\sigma^X_{\rm pix}$  (and $\sigma^P_{\rm pix}=\sqrt{2}\sigma^T_{\rm pix}$)
    depends on the instrumental parameters;
   $\theta_{\rm fwhm}$ is the full width at half maximum (FHWM) of the Gaussian beam.
\\

 \begin{figure}
\begin{center}
\includegraphics[trim = 1mm 1mm 2mm 1mm, clip, width=11 cm, height=5cm]{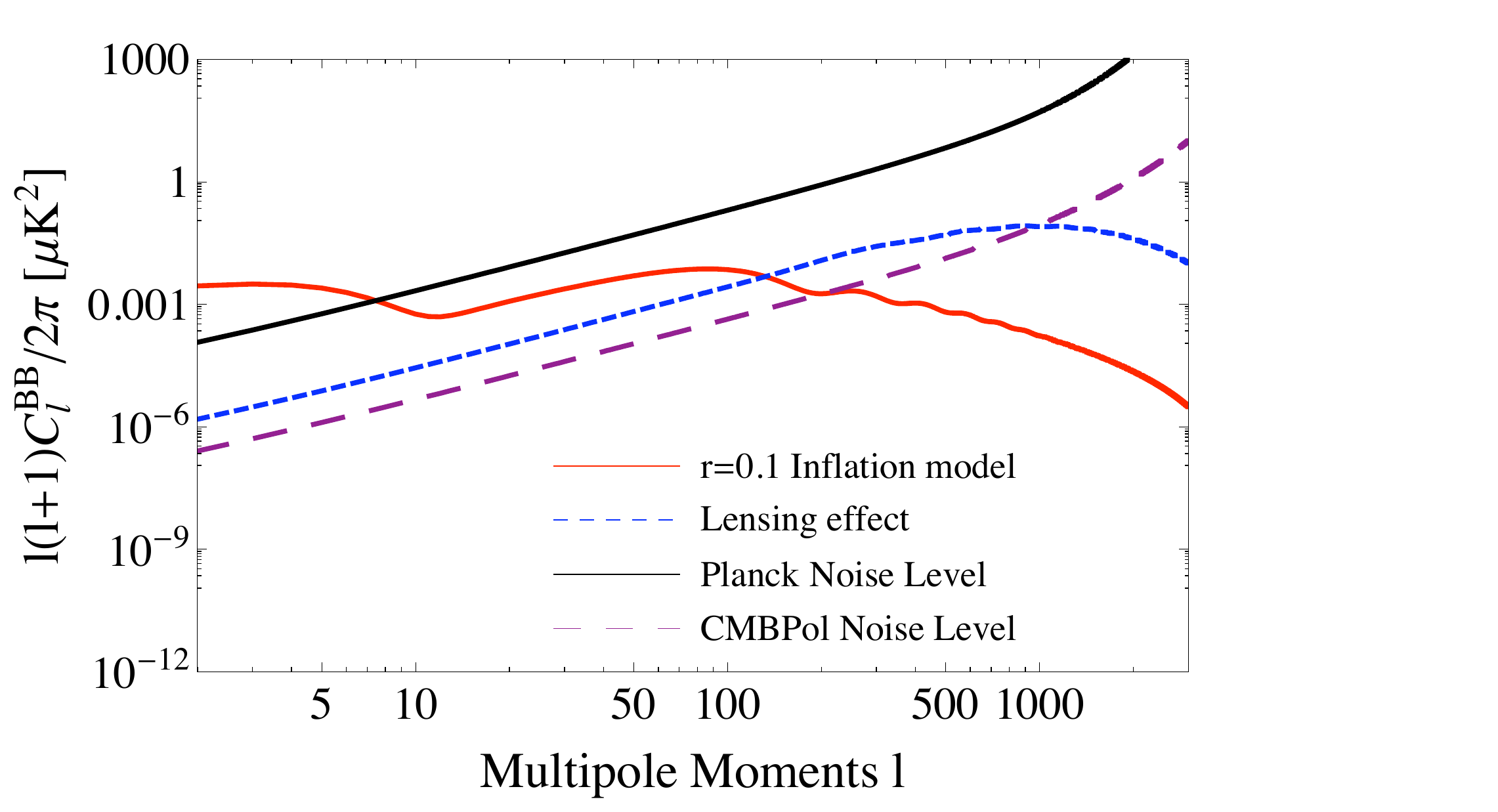}
\caption{Polarization noise power spectra for forthcoming
  experiments. Note that these curves include uncertainties associated
  with the instrumental beam. The red line shows the $B$-mode power
  spectrum for the standard inflationary model with $r=0.1$.} \label{fig:SN}
\end{center}
\end{figure}

\noindent For the Planck experiment, we include three
channels with frequencies (100 GHz, 143 GHz, 217 GHz) and noise
levels per beam $(\sigma^T_{\rm pix})^2$= (46.25 $\mu{\rm K}^2$,
36 $\mu{\rm K}^2$, 171  $\mu{\rm K}^2$). The FHWM of the three
channels are $\theta_{\rm fwhm}$ =($9.5, 7.1, 5.0$) arc-minute. 
These figures are taken from the values given in \cite{Planck}.
We combine three channels for the CMBPol experiment \cite{CMBPol}
with frequencies (100 GHz, 150 GHz, 220 GHz), noise levels
$(\sigma^T_{\rm pix})^2$ = (729 nK$^2$, 676 nK$^2$, 1600 nK$^2$)
and $\theta_{\rm fwhm}$ = ($8, 5, 3.5$) arc-minute. Sky coverages
of $f_{\rm sky}$ = $0.65, 0.8$ are respectively assumed and integration time of 14 months.
In Figure~\ref{fig:SN} we show the noise levels for these experiments as
a function of multipole number $\ell$. The blue line corresponds to the B-mode power spectrum
 using the standard power-law parameterisation with $r=0.1$.
 The lensed $C^B_\ell$ is also shown in the same Figure, which can be treated as a part 
 of the total noise power spectrum $N^B_\ell$ as well as the instrumental noise power
 spectra \cite{Perotto06}. For more information of the noise and beam profile of
 each frequency channel please refer to \cite{Yin10}.

\section{Primordial power spectra constraints}
\label{sec:spec}

\subsection{Power-law parameterisation}
\label{sec:ns}

\noindent
Because slow-roll inflation predicts the spectrum of curvature perturbations
to be close to scale-invariant, the simplest proposal is to assume that the 
initial spectrum has a power-law form, parameterised by
\begin{equation}
  \mathcal{P_R}(k)=A_{\rm s} \left( \frac{k}{k_0}  \right)^{n_{\rm s}-1},
\end{equation}
where the {\it spectral index} $n_{\rm s}$ is expected to be close to
unity.
A spectrum where the typical amplitude of perturbations is identical
on all length scales is known as Harrison-Zel'dovich spectrum ($n_{\rm s}=1$),
and it has been ruled out by several studies (see for instance \cite{Vazquez12}).
Here, we assume, for simplicity, that the tensor spectrum
is also described by a  power-law function:
\begin{equation}
  \mathcal{P_T}(k) = A_{\rm t} \left( \frac{k}{k_0} \right)^{n_{\rm t}},
\end{equation}

\noindent
 where the tensor amplitude $A_t$ is related to tensor-to-scalar ratio $r_{\rm s}=A_t/A_s$.
 For this parameterisation we assume that $r(k_0)$ and the tensor spectral 
 index $n_{\rm t}\equiv d\ln \mathcal{P_T}(k)/d\ln k$
 satisfy the consistency relation for a single field slow-roll inflation
 $n_{\rm t}= -r_{\rm s}/8$ \cite{Marina11}.
The  power-law parameterisation thus contains only three free parameters:
$A_{\rm s}$, $n_{\rm s}$, and $r_{\rm s}$. For these parameters, we assume a prior
$A_{\rm s}=[1,50]\times10^{-10}$ for the amplitude,  a conservative prior for the
spectral index $n_{\rm s}=[0.7,1.2]$ and a tensor-to-scalar ratio prior of $r_{\rm s}=[0,1]$.
\\

\begin{figure}
\begin{center}
$(n_s)$\\ 
$\begin{array}{cc}
\includegraphics[trim = 25mm 80mm 25mm 80mm, clip, width=7 cm, height=6.0cm]{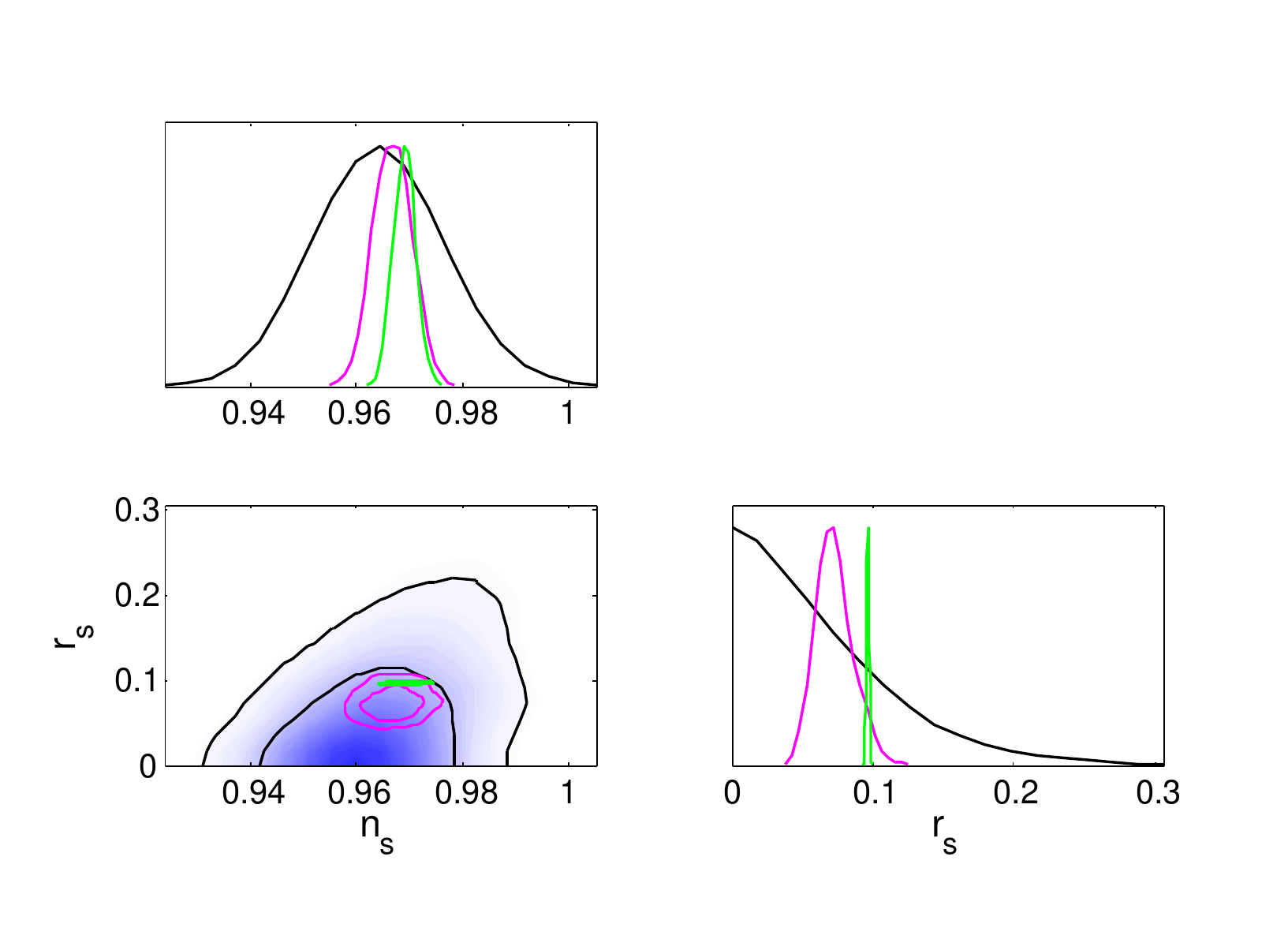}&
\includegraphics[trim = 1mm -10mm 1mm 10mm, clip, width=8. cm, height=5.cm]{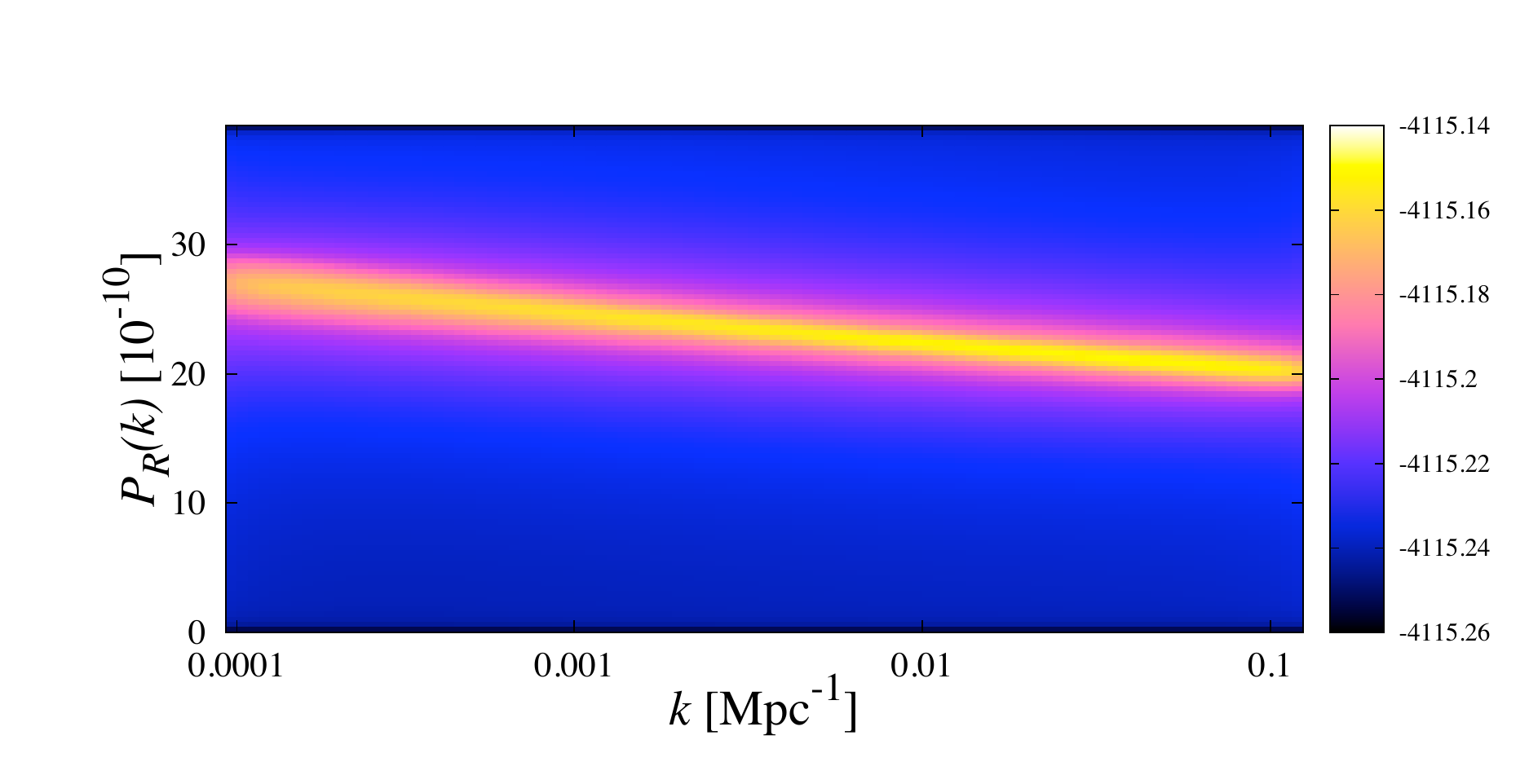}
\end{array}$
  \scalebox{0.95}{
\begin{tabular}{cccc}
\cline{1-4}\noalign{\smallskip}
\hline
\vspace{0.0cm}
&\quad Current  & \quad Planck  &\quad CMBPol     \\
&\quad   & \quad  ($\sigma_i$) &\quad ($\sigma_i$)
\\ \hline
\vspace{0.1cm}
$A_{\rm  s}[10^{-10}]$ &        $22.74\pm 0.63$ &       $ 0.19$	 &0.131          \\
\vspace{0.1cm}
$n_{\rm s}$ &            		   $0.964\pm0.011$&        $ 0.003$    &0.002  \\
\vspace{0.1cm}
$r_{\rm s}$ &              		     $<0.171$ &		          $0.013$	&0.0009      \\
\hline
\hline
\end{tabular}
}
\end{center}
\caption{
Left panel: 1D and  2D probability posterior distributions for the  power spectrum parameters,
assuming a simple tilt parameterisation ($n_{\rm s}$); using both current observations (black line)
 and future experiments (red for Planck and green for CMBPol).  2D constraints are plotted  
 with $1\sigma$ and $2\sigma$ confidence contours. Right panel: Reconstruction of the scalar 
 spectrum using present data; lighter regions represent an improved fit.
}
\label{fig:ns}
\end{figure}

Figure~\ref{fig:ns} shows 1D and 2D  marginalised posterior distributions of the
scalar spectrum index $n_{\rm s}$ and the tensor-to-scalar ratio $r_{\rm s}$, using both 
current cosmological observations (black line) and future experiments 
(red for Planck and green for CMBPol).
 The bottom panel shows the limits imposed by
current and future experiments. For present observations: $n_{\rm
s}=0.964\pm0.011$ and $r_{\rm s}<0.171$ (mean values of  68\% C.L.
are quoted for two-tailed distributions, whilst one-tailed
distribution only the upper 95\% C.L.). These results are in
agreement with previous studies, i.e. \cite{WMAP,ACT,SPT}. With
regards to future constraints, we have used mean values obtained
from current observations as the fiducial model (with fixed
$r_{\rm s}=0.1$). We  notice that $1\sigma$ error bars of the
spectral index $n_{\rm s}$, shown in the bottom panel of the same
Figure, reduce by about four times using a Planck-like experiment
and five times for a CMBPol experiment. Whereas Planck will be able
to distinguish tensor components  with an accuracy of
$\sigma_r=0.013$, this is highly improved by CMBPol data $\sigma_r=0.0009$.
If we consider only one channel for comparison, e.g. 100 GHz, the constraints
on the tensor-to-scalar ratio are given by $\sigma_r=0.02$, 
in agreement with previous results \cite{Burigana10}.
The top-right panel of Figure~\ref{fig:ns} illustrates the resulting shape of
$\mathcal{P_R}(k)$ corresponding to the posterior  distributions
using present data.

\subsection{Running scalar spectral-index}
\label{sec:run}

A further extension is possible by allowing the scalar spectral index to vary
as a function of scale, such that $n_{\rm s}(k)$.  This can be achieved by
including a second order term in the expansion of the power spectrum
\begin{equation}
 \mathcal{P_R}(k) = A_{\rm s} \left( \frac{k}{k_0} \right)^{n_{\rm s}-1+(1/2)\ln (k/k_0)(dn/d\ln k)},
\end{equation}
where $n_{\rm run} \equiv dn_{\rm s}/d \ln k$ is termed the {\it running of the tilt}
 and we would expect $n_{\rm run} \approx 0$ for standard
inflationary models. We have kept the same tensor spectrum as in the simple power-law 
parameterisation, with a tensor-to-scalar ratio $r_{\rm run}$ at a scale of  $k_0= 0.015$ 
$\rm{Mpc}^{-1}$ to avoid correlations amongst parameters \cite{Marina07}. 
We maintained the same priors for the inflationary parameters $A_{\rm s}$, $n_{\rm s}$, 
and $r_{\rm run}$ and select a prior of the running parameter of $n_{\rm run}=[-0.1,0.1]$ 
as used by \cite{Parkinson10}.
\\

\begin{figure}
\begin{center}
$(n_{\rm run}) \,\, \mathcal{B}_{n_{\rm run},n_{\rm s}} =+2.0 \pm 0.3$\\
$\begin{array}{cc}
\includegraphics[trim = 20mm 80mm 10mm 80mm, clip, width=9.5 cm, height=6.5cm]{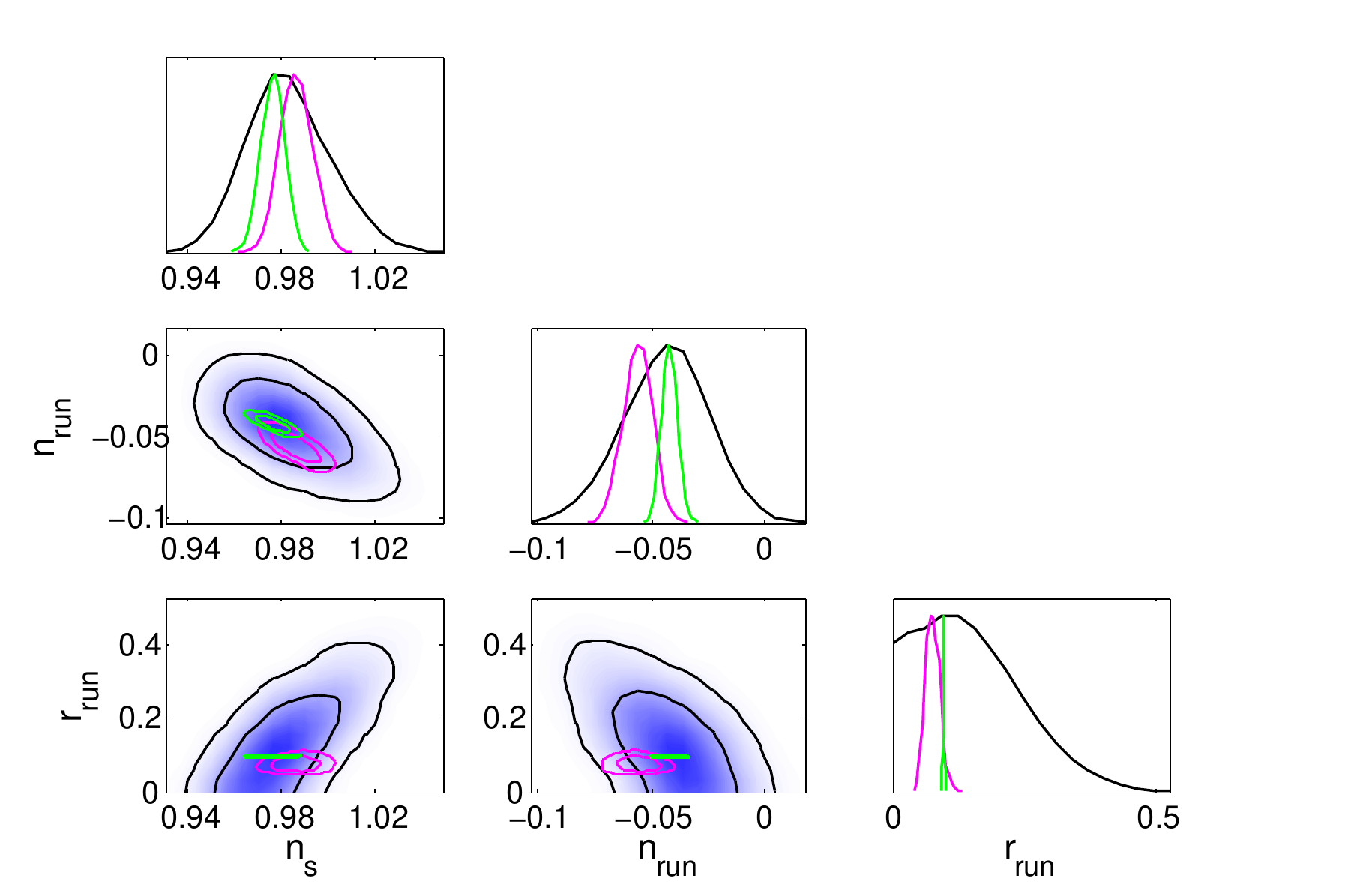}&
\includegraphics[trim = 12mm -10mm -20mm 10mm, clip, width=8. cm, height=5.5cm]{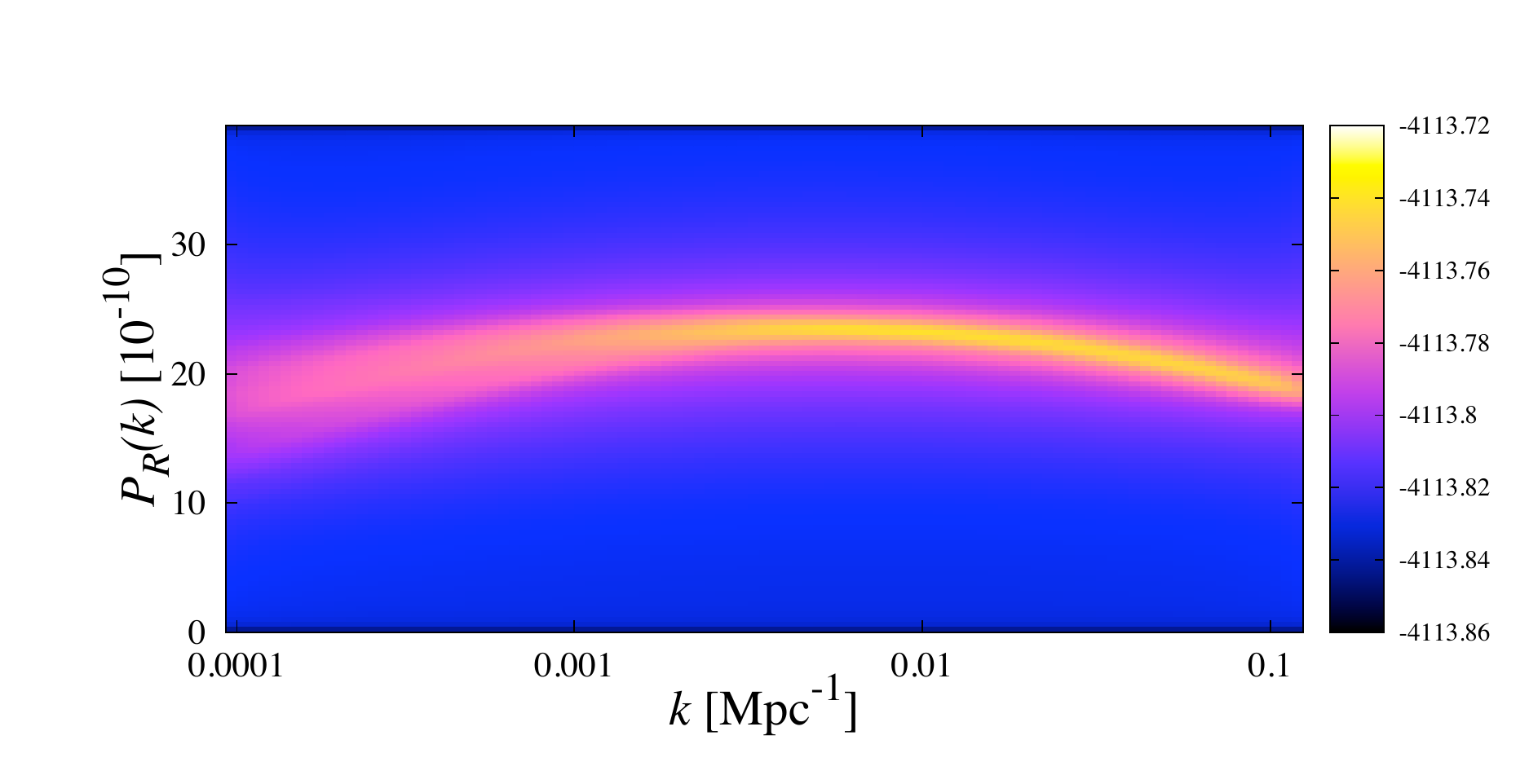}
\end{array}$
  \scalebox{0.95}{
\begin{tabular}{cccc}
\cline{1-4}\noalign{\smallskip}
\hline
\vspace{0.0cm}
&\quad Current  & \quad Planck  &\quad CMBPol     \\
&\quad   & \quad  ($\sigma_i$) &\quad ($\sigma_i$)
\\ \hline
\vspace{0.1cm}
$A_{\rm  s}[10^{-10}]$ &        $23.66\pm 0.83$ &       $ 0.19$ &0.0142         \\
\vspace{0.1cm}
$n_{\rm s}$ &               $0.985\pm0.017$&        $ 0.0064$    &0.0048 \\
\vspace{0.1cm}
$n_{\rm run}$ &         $-0.043\pm0.018$&       $ 0.0062$    & 0.0034    \\
\vspace{0.1cm}
$r_{\rm run}$ &                 $<0.324$ &          $0.013$&0.001     \\
\hline
\hline
\end{tabular}
}

\end{center}
\caption{
Left panel: 1D and  2D probability posterior distributions for the inflationary parameters,
assuming a power-law with a running parameter ($n_{\rm run}$);
using both current data (black line) and future experiments (red for Planck and green for CMBPol).
 2D constraints are plotted  with $1\sigma$ and $2\sigma$ confidence contours.
Right panel: Reconstruction of the scalar spectrum using present data; lighter regions represent 
an improved fit. The top label  denotes the Bayes factor of the $n_{\rm run}$-model compared
  to the power-law $n_{\rm s}$-model, using current observations.}
\label{fig:nrun}
\end{figure}

Figure \ref{fig:nrun} shows the 1D and 2D marginalised posterior distributions for the 
inflationary parameters, using current experiments (black line): $n_{\rm s}=0.985\pm0.017$, $n_{\rm
run}=-0.043\pm0.018$ and $r_{\rm run}<0.324$; and Planck (red line) and CMBPol (green line) 
realisations. The top label of the figure indicates the Bayes factor using present observations, which
in this case and throughout the paper is compared with respect to the power-law parameterisation.
We first note that in the presence of a tensor component the bending of the scalar
spectrum is enhanced through a larger running parameter.
That is because, at the largest scales, the contribution of the CMB-tensor 
spectrum compensates the power of the CMB-scalar, leaving hence the total 
CMB spectrum unaffected; for instance
$n_{\rm run}=-0.043\pm0.018$ compared to $n_{\rm run}=-0.028\pm0.014$ without tensors.
We also observe that using current experiments a negative $n_{\rm run}$
parameter in preferred by more than $2.5\sigma$~C.L.
Hence the necessity to include a turn-over in the power spectrum.
 This result is confirmed by noticing the Bayes factor is significantly favoured compared 
 to the simple power-law model, $\mathcal{B}_{n_{\rm run},n_{\rm s}} =+2.0 \pm 0.3$.
 Considerations of the running of running of the spectral index are also being explored \cite{Powell12}.
  We notice that  correlations created by the inclusion 
 of the running parameter broaden the constraints
on the tensor-to-scalar ratio by about 1.5 times. Future constraints are also broadened compared
to the power-law parameterisation. The summary of the constraints on the inflationary parameters
 is shown in the bottom panel of Figure~\ref{fig:nrun}, and the reconstruction of $\mathcal{P_R}(k)$,
 using present data, in the top-right panel.

\subsection{Model independent reconstruction}
\label{sec:2ki}

We have seen that deviations from the simple power-law, by the introduction of the running 
parameter, are relevant in explaining present data.
In order to corroborate this result and look for deviations from the power-law parameterisation, 
we consider a model-independent reconstruction. The reconstruction process we follow is 
based on the approach used previously by \cite{Vazquez12, Vazquez13}. 
We place two fixed $k$-nodes at
sufficiently separated positions $[k_{\rm min},k_{\rm max}]$, with varying amplitudes
[$A_{{\rm s},k_{\rm  min}}$, $A_{{\rm s},k_{\rm  max}}$], and place inside
additional `nodes' with the freedom to move around in both position
$k_i$ and amplitude $A_{{\rm s},k_i}$. We assume that most of the astrophysical
information is encompassed within the scales $k_{\rm min} =0.0001$ ${\rm Mpc}^{-1}$ 
and $k_{\rm max}=0.3$ ${\rm Mpc}^{-1}$.
Outside of these limits we consider the spectrum to be constant with values
equal to those at $k_{\rm min}$ and $k_{\rm max}$ respectively.
 We allow variations in the amplitudes  with a conservative prior $A_{{\rm s},k_i}\in [1,50] \times 10^{-10} $.
To maintain continuity between $k$-nodes, a linear  interpolation is performed such that
the form of the power spectrum is described by
\begin{eqnarray} 
\mathcal{P_R}(k) = \left\{ \begin{array}{ll} 

A_{{\rm s},k_{\rm min}} 				& \quad k\le k_{\rm min},\\ 
A_{{\rm s},k_i }			      		& \quad	k_{\rm min}< k_i<k_{i+1}< k_{\rm max},   \\ 
A_{{\rm s},k_{\rm max}}	 			&\quad 	k\ge k_{\rm max},
\end{array} \right. && \\ \nonumber \\ 
{\rm and\,\, with\,\,  linear\,\,interpolation\,\, for \quad} 
  k_{\rm min}\le &k_i&\le k_{\rm max}. \nonumber 
\end{eqnarray}



\noindent
 We have  restricted the model-independent reconstruction to two internal-nodes  which
 we consider are sufficient to provide an accurate description of the shape of the power spectrum.
The tensor spectrum is parameterised by a power-law form, similarly to the one in Section \ref{sec:ns}.
Here the tensor-to-scalar ratio, given by $r_{2k_i}=A_t/P_{\mathcal R}(k_0)$, is  computed at the
scale $k_0=0.015$ Mpc$^{-1}$ and also satisfies the consistency relation
$r_{2k_i}=-n_t/8$; with prior $r_{2k_i}=[0,1]$.
\\

\begin{figure}
\begin{center}
$(2{\rm k_i}) \,\, \mathcal{B}_{2{\rm k_i},n_{\rm s}} =+2.3 \pm 0.3$\\
 \vspace{0.5cm}

  \begin{minipage}[l]{0.5\textwidth}
$\begin{array}{cc}
\includegraphics[trim = 35mm 70mm 1mm 70mm, clip, width=9.5cm, height=8.5cm]{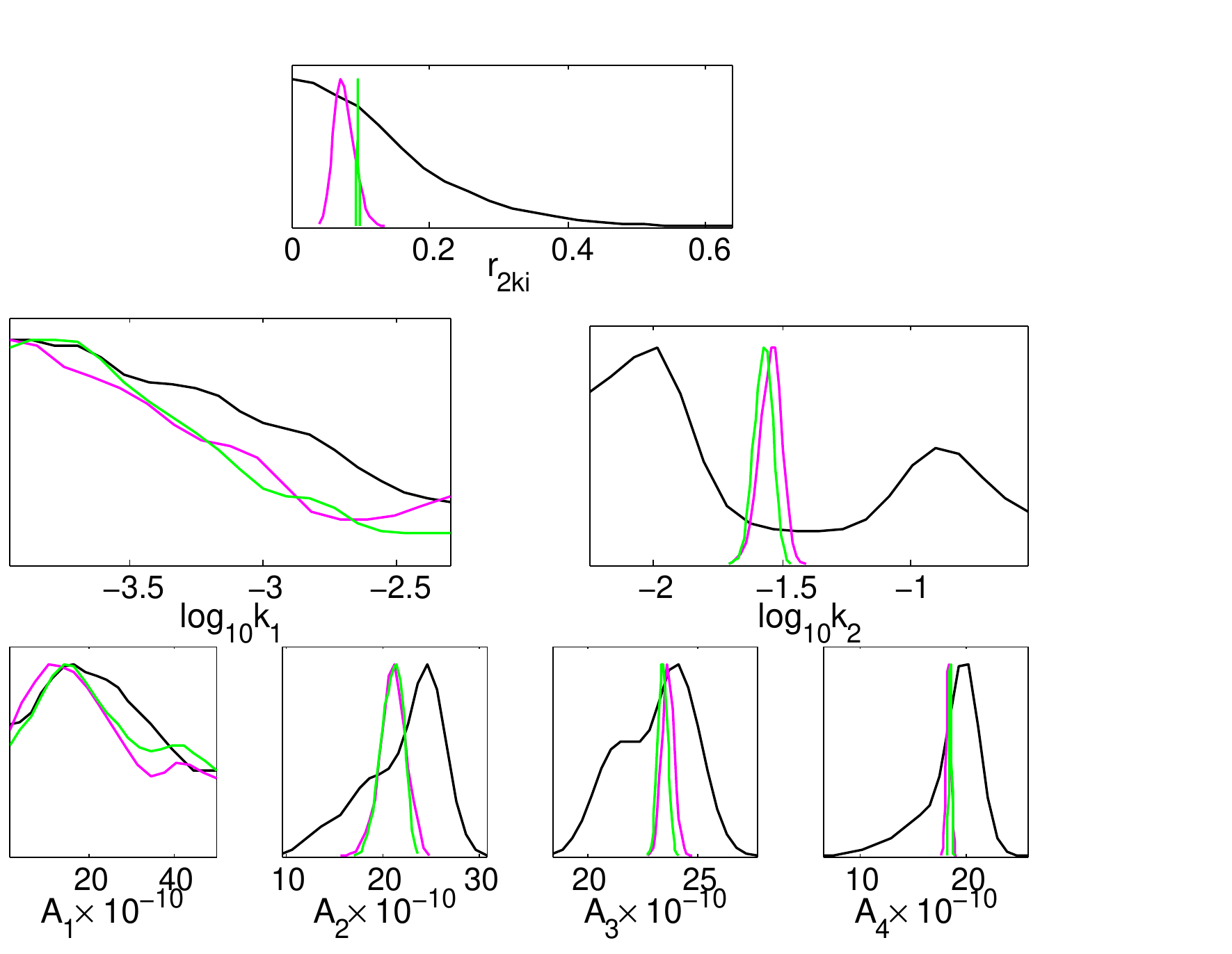}
\end{array}$
  \end{minipage}
  \begin{minipage}[r]{0.45\textwidth}
\qquad  $\begin{array}{c}
\includegraphics[trim = 1mm 1mm 1mm 10mm, clip, width=8. cm, height=4.3cm]{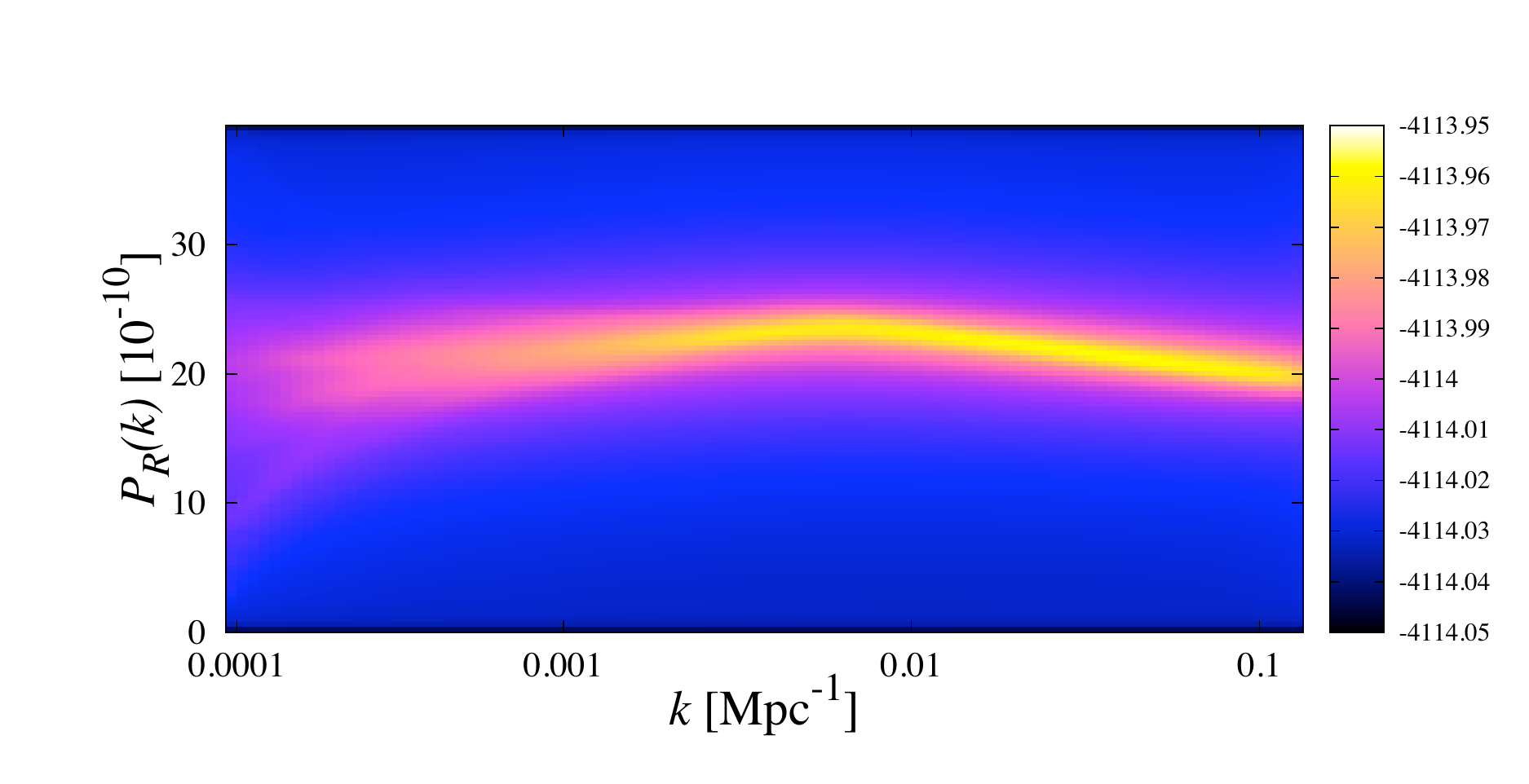}\\
\scalebox{0.85}{
\begin{tabular}{cccc}
\cline{1-4}\noalign{\smallskip}
\hline
\vspace{0.0cm}
&\quad Current  & \quad Planck  &\quad CMBPol     \\
&\quad   & \quad  ($\sigma_i$) &\quad ($\sigma_i$) \\
 \hline
\vspace{0.1cm}
$A_{\rm  1}[10^{-10}]$ &            $-$ &       $ - $ & $-$         \\
\vspace{0.1cm}
$A_{\rm 2}[10^{-10}]$ &         $22.02^{+3.74}_{-4.43}$&        $ 1.29$ &1.09   \\
\vspace{0.1cm}
$A_{\rm 3}[10^{-10}]$ &         $23.68^{+1.72}_{-1.98}$&        $ 0.27$ &0.21   \\
\vspace{0.1cm}
$A_{\rm 4}[10^{-10}]$ &         $18.67^{+2.06}_{-2.37}$&        $ 0.20$ &0.10   \\
\vspace{0.1cm}
$\log_{10}k_{\rm 1}$ &          $-$&        $ -$    &-   \\
\vspace{0.1cm}
$\log_{10}k_{\rm 2}$ &          $-$&        $ 0.04$ &0.034  \\
\vspace{0.1cm}
$r_{2ki}$ &                     $<0.34$&                $ 0.014$    &0.00096    \\
\hline
\hline
\end{tabular}
}
\end{array}$
  \end{minipage}

\end{center}
\caption{
Left panel: 1D and  2D probability posterior distributions for the  power spectrum parameters,
assuming a two internal-node reconstruction ($2k_i$); using both current cosmological 
observations (black line) and future experiments (red for Planck and green for CMBPol).
 2D constraints are plotted  with $1\sigma$ and $2\sigma$ confidence contours.
 Right panel: Reconstruction of the scalar spectrum using present data;
 lighter regions represents an improved fit. Top label  denotes the Bayes factor of the 
 $2k_i$-model compared to the power-law $n_{\rm s}$-model, using current observations.}
\label{fig:2ki}
\end{figure}

The top-left panel of Figure \ref{fig:2ki} displays the 1D and 2D marginalised posterior
distributions for the parameters used in the model-independent reconstruction.
At the largest scales, we observe the lack of tight
constraints on the amplitude $A_1$, mainly due to
the cosmic variance and correlations with other parameters.
At smaller scales, the constraints on the amplitues (i.e. $A_2$,  $A_3$ and $A_4$) get tighter.
We notice the presence of a bi-modal distribution in the medium/small scales,  represented by $k_2$,
 where  the highest peak ($k\sim 0.01 $Mpc$^{-1}$)
matches the position of the turn-over in the primordial spectrum, as  seen in the top-right panel  of
Figure~\ref{fig:2ki}.
The other peak is located where the constraints seem  to improve by updated data sets:
at the overlapping of WMAP/ACT observations ($0.1<k<0.14$) with  LRG7 measurements.
The reconstructed spectrum clearly presents a turn-over, however with the bending at small 
scales less pronounced that in the running model.
Notice that the Bayes factor, shown in the top label of the same Figure,
is significantly preferred over the simple tilt model, even though the
two-internal-node reconstruction contains four additional parameters; it is also marginally 
preferred over the running model.
Future experiments will be able to pin-down accurately the shape of the primordial spectrum at medium
and small scales ($k_2$), however at the largest scales ($k_1$) the cosmic variance still dominates,
as seen in the 1D posterior distribution of $A_1$.
Current and  future constraints of the inflationary parameters are summarised in the bottom 
panel of Figure \ref{fig:2ki}.

\subsection{Lasenby \& Doran model}
\label{sub_sec:LD}

The Lasenby \& Doran model is based on the restriction of the total conformal time
available in a closed universe \cite{Lasenby05}.
At the largest scales, the predicted scalar and tensor spectra naturally incorporate a drop-off 
without the need to parameterise them, whilst, at small scales they mimic a slight running behaviour.
An important point to bear in mind, is that in the LD model 
the functional forms of $H$  and $\dot \phi$ during inflation are expressed
 using just two parameters $b_0$ and $b_4$ \cite{Vazquez11,Vazquez12}. These parameters describe the 
 initial conditions, along with the  standard cosmological parameters, and therefore the primordial 
 spectra generated by the LD model are given in terms of
 \begin{equation}
P_{\mathcal R}(k)=P_{\mathcal R}(k; b_0,b_4,\Omega_i,H_0),\qquad
P_{\mathcal T}(k)=P_{\mathcal T}(k; b_0,b_4,\Omega_i,H_0).
 \end{equation}

 \noindent
Notice that the tensor-to-scalar ratio $r_{\rm LD}$ is a derived quantity
in terms of the cosmological parameters, $H_0$, $\Omega_i$, and the initial-conditions 
parameters $b_0$ and $b_4$:
\begin{equation}
r_{LD}(k)=r_{LD}(k; b_0,b_4,\Omega_i,H_0).
\end{equation}

\begin{figure}
\begin{center}
\includegraphics[trim = 2mm 1mm 1mm -1mm, clip, width=5 cm, height=3.5cm]{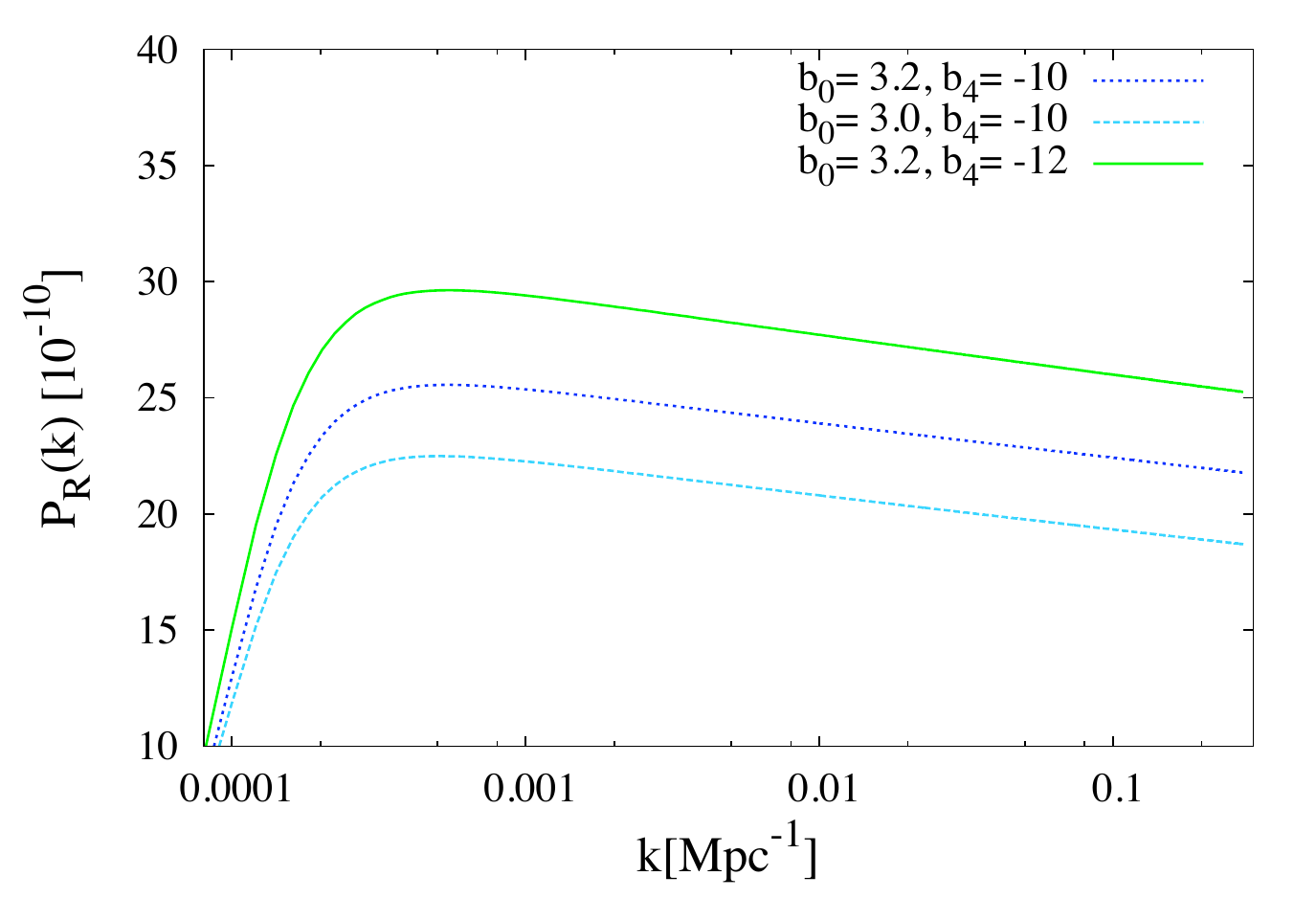} 
\includegraphics[trim = 2mm 1mm 1mm -1mm, clip, width=5cm, height=3.5cm]{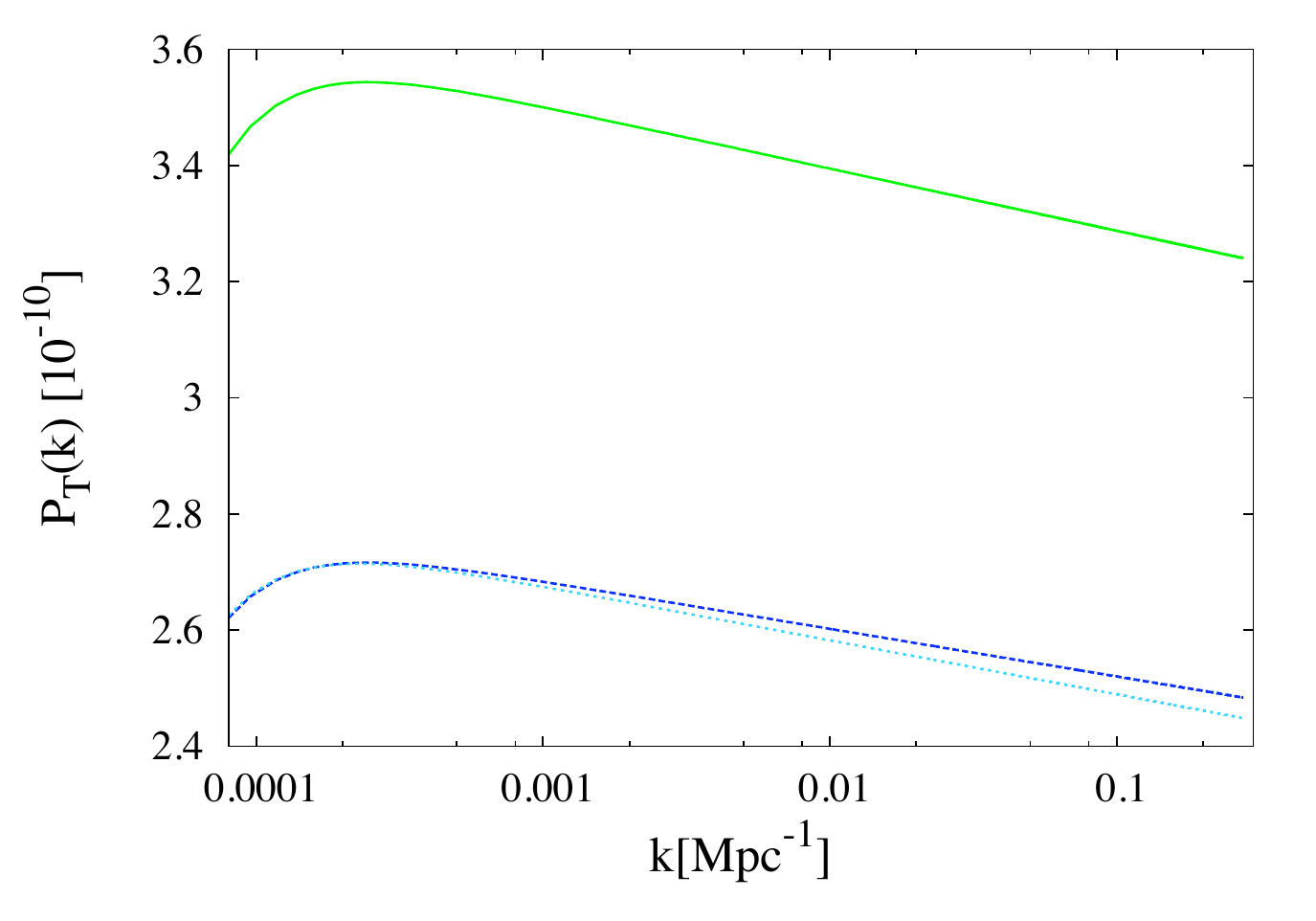}
\includegraphics[trim = 1mm 1mm 5mm 1mm, clip, width=5. cm, height=3.5cm]{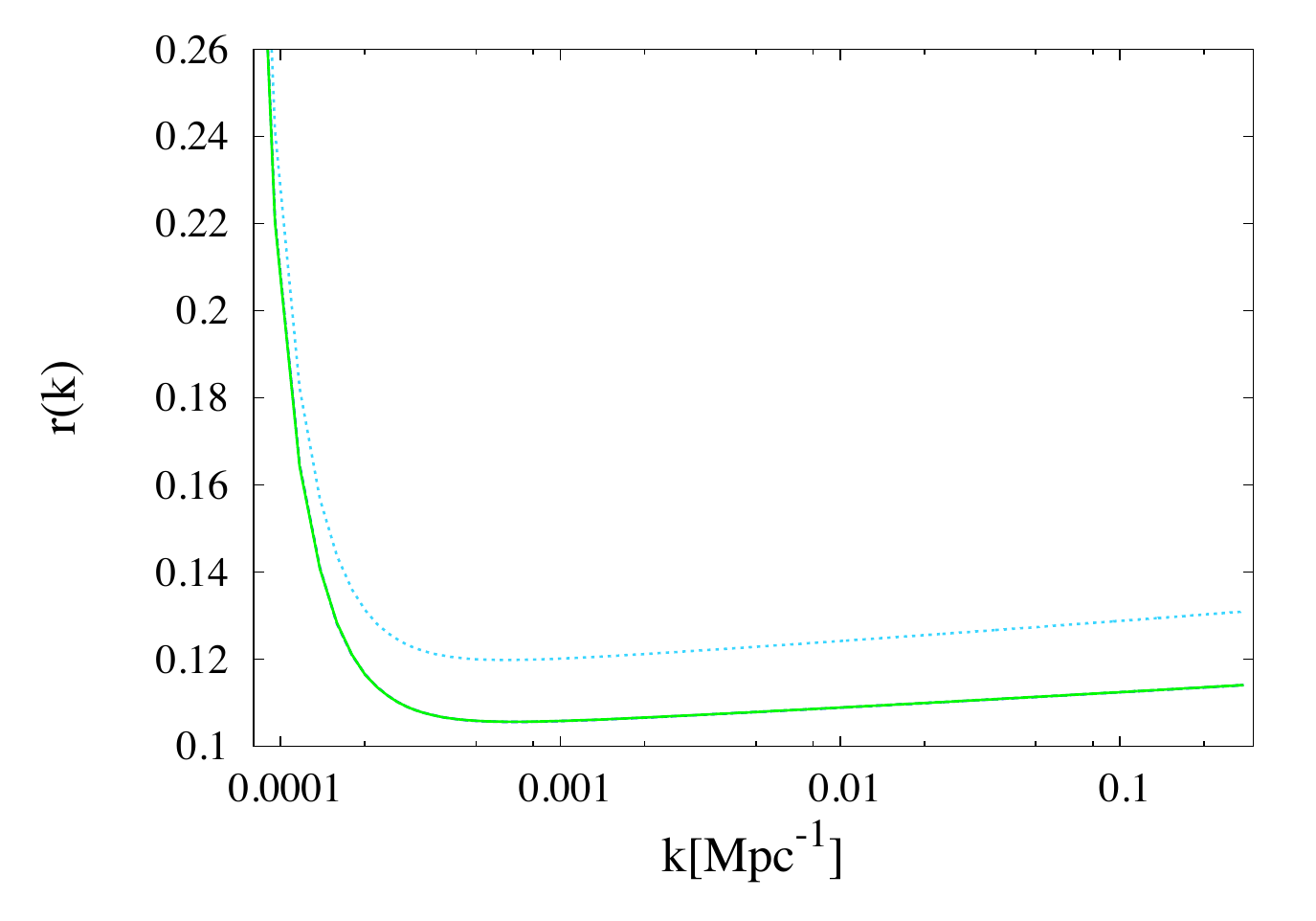}
\end{center}
\caption{Derived primordial power spectra, $P_{\mathcal{R}}(k)$ and $P_{\mathcal{T}}(k)$,
 from the Lasenby \& Doran model, using a different set of parameters of $b_0$ and $b_4$ (left and
 middle panels);
 units of $b_4$ are given in $[10^{-10}]$.
 The right panel  shows the tensor-to-scalar ratio $r_{\rm LD}=P_{\mathcal T}/P_{\mathcal R}$.}
\label{fig:LD_r}
\end{figure}

\begin{figure}
\begin{center}
$({\rm LD}) \,\, \mathcal{B}_{{\rm LD},n_{\rm s}} =+3.4 \pm 0.3$\\
$\begin{array}{cc}
\includegraphics[trim = 20mm 85mm 20mm 90mm, clip, width=8.5 cm, height=6.cm]{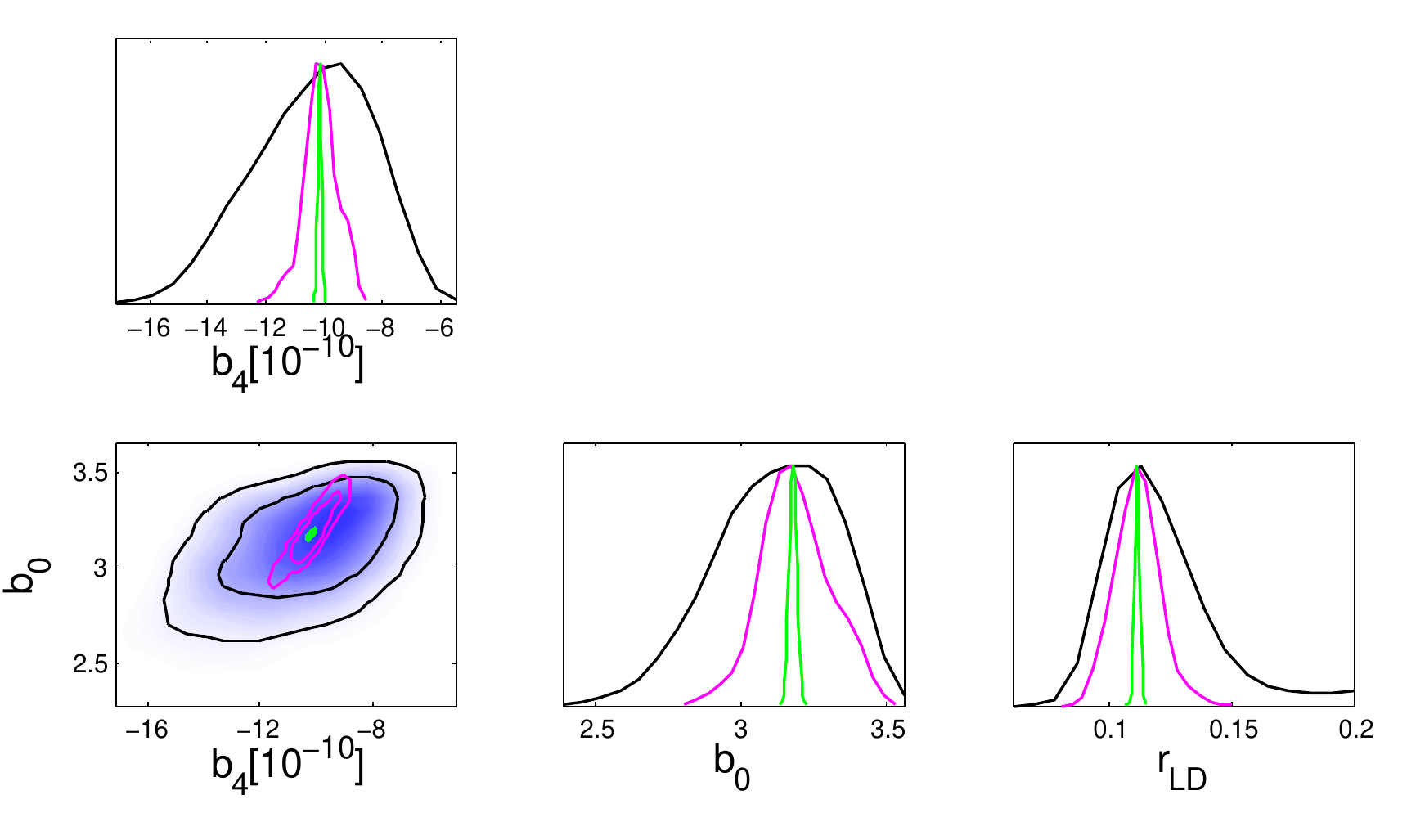}
\includegraphics[trim = 1mm 1mm 1mm 10mm, clip, width=8. cm, height=4.5cm]{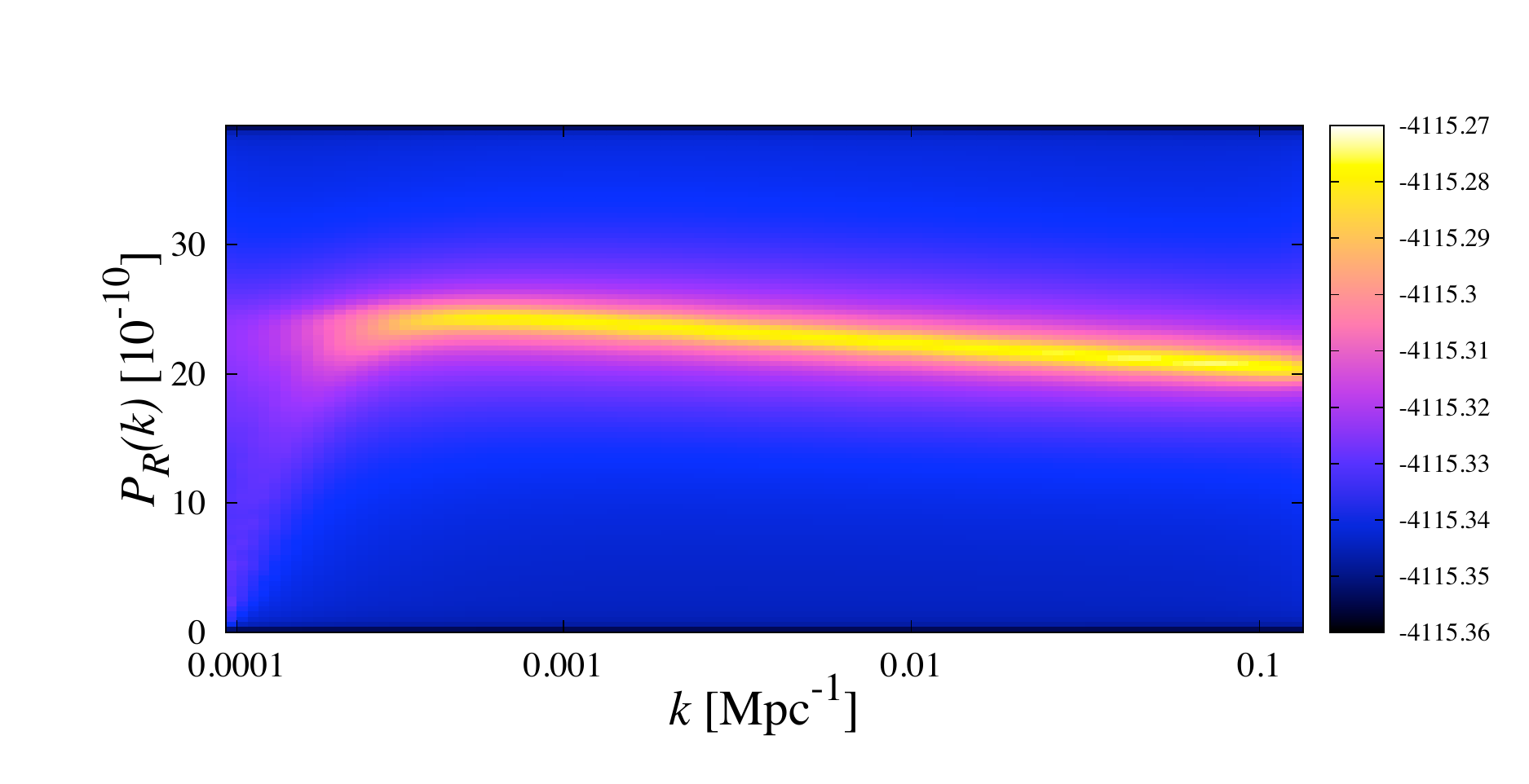}
\end{array}$
  \scalebox{.95}{
\begin{tabular}{cccc}
\cline{1-4}\noalign{\smallskip}
\hline
\vspace{0.0cm}
&\quad Current  & \quad Planck  &\quad CMBPol     \\
&\quad   & \quad  ($\sigma_i$) &\quad ($\sigma_i$)
\\ \hline
\vspace{0.1cm}
$b_{\rm  0}$ &              $3.21\pm0.41$ & $ 0.24$ & $0.013$           \\
\vspace{0.1cm}
$b_{\rm 4}[10^{-10}]$ &     $-10.39\pm 2.95$&    $ 1.09$    &$0.065$    \\
\vspace{0.1cm}
$^{*}\,\,r_{\rm LD} $ &             $0.11\pm{0.024}$ &  ${0.017}$&$0.002$       \\
\hline
\hline
\end{tabular} 
}

\end{center}
\caption{1D and  2D probability posterior distributions for the  power spectrum parameters,
assuming a Lasenby \& Doran model (LD); using both current cosmological observations (black line)
 and future experiments (red for Planck and green for CMBPol).
 2D constraints are plotted  with $1\sigma$ and $2\sigma$ confidence contours.
 The top label   denotes the associated Bayes factor with  respect to the power-law 
 $n_{\rm s}$ model using present  data. $^*$In this model, $r_{\rm LD}$ is a derived parameter.}
\label{fig:LD}
\end{figure}

\noindent
That is, if we use  values of $b_0$ and $b_4$ along with the cosmological parameters there is  
no need to introduce  additional variables to describe the tensor-to-scalar ratio $r_{LD}$.
Figure~\ref{fig:LD_r} shows the primordial spectra, both scalar and tensor, for a given combination 
of $b_0$ and $b_4$ parameters. In the right panel of this Figure, we illustrate  the tensor-to-scalar 
ratio and its  degeneracy with a selection of the parameters, for instance, the combination of 
$\{b_0=3.2, b_4=-10\times 10^{-10} \}$ or $\{b_0=3.0, b_4=-12\times 10^{-10} \}$ provides the 
same tensor-to-scalar ratio, even though their scalar and tensor  spectra differ considerably.
For  further details about the LD model see, for instance \cite{Lasenby04, Lasenby05, Vazquez11}.
 To compute the LD spectra we refer to \cite{Vazquez11,Vazquez12}. We have also 
 chosen the priors based on the same paper: $\Omega_k =[-0.05, 10^{-4}]$, $b_0 =[1,4] $,  
 $b_4=[-30,-1]\times 10^{-9}$. Figure~\ref{fig:LD} shows 1D and 2D marginalised posterior 
 distributions of the parameters involved in the description of the LD model.
 A novel result from the LD model is that its constraints on the tensor-to-scalar ratio are 
 different from zero: $r_{\rm LD}=0.11 \pm 0.024$, contrary to the standard power-law parameterisation.
This happens mainly due to the $\phi^2$-type potential assumed in the model.  
 The Bayes factor of this  model compared to the simple-tilt  
 model is shown in the  top label of the same Figure. The low number of parameters and the 
 reduced power at both large and small scales make the LD model strongly favoured compared to the simple 
 tilt and significantly so compared to the running and the  two-internal-node reconstruction.
Future experiments will provide an insight on discriminating amongst models, as we  will see in the
next section.

\section{Model Selection}
\label{sec:mod_sel}

Throughout the analysis,
we have included the Bayes factor for each model and found that the Lasenby \& Doran model is the most
preferred by current observations. Future experiments will be able to  distinguish between 
models more effectively. Let us assume for a moment that the LD spectra represent the  true model.
We then use the LD spectra, with best-fit values obtained by using present data 
(shown in the bottom panel of Figure \ref{fig:LD}), as the fiducial model to simulate future CMB 
observations.  We analyse this mock data to reconstruct the input spectrum using the
set of models aforementioned.
Table \ref{tab:ms} shows the Bayes factor for the different parameterisations 
compared to the LD model,
along with the recovered tensor-to-scalar ratio.
There is indeed a distinction between models, with the data clearly indicating a preference 
for the LD model, used to generate the input-simulated data. Idealised Planck results might provide decisive 
conclusions on distinguishing the LD model from the simple-tilt $n_{\rm s}$-model, 
$\mathcal{B}_{n_s,LD}=-6.3\pm 0.3$, and a running $n_{\rm run}$-model,  
$\mathcal{B}_{n_{\rm run},LD}=-6.5\pm 0.3$, and strong preference when compared to the  
two-internal-node reconstruction $2k_i$-model, $\mathcal{B}_{2k_i,LD}=-3.1\pm 0.3$.
There will also be a strong  preference for the model independent reconstruction over both 
the $n_{\rm s}$ and $n_{\rm run}$ models: $\mathcal{B}_{2k_i,n_{\rm s}}=+3.2\pm 0.3$ and  
$\mathcal{B}_{2k_i,n_{\rm run}}=+3.4\pm 0.3$, respectively.
With regards to the CMBPol experiment, this might definitely differentiate the LD spectrum from 
the rest of the spectra. In contrast to the Planck experiment, the model-independent reconstruction 
for CMBPol is now strongly  favoured compared to the $n_{\rm run}$ model.
CMBPol also provides a strong preference to differentiate the simple-tilt model $n_{\rm s}$ over the
running model $n_{\rm run}$: $\mathcal{B}_{n_{\rm s},n_{\rm run}}=+2.5\pm 0.3$.
Therefore, future experiments certainly will be able to differentiate between these models and pin down
the right form of the primordial spectrum.

\newcolumntype{C}[2]{>{\hsize=#1\hsize\columncolor{#2}\centering\arraybackslash}X}

  \begin{table}
\begin{center}
\caption{Model Selection. The input spectrum, given by the LD model, is reconstructed using
 different models.
We  show the Bayes factor for  each model $\mathcal{B}_{i,LD}$,
along with the recovered tensor-to-scalar ratio $r_i$.  }
\begin{tabular}{ccccc}
\cline{1-5}\noalign{\smallskip}
\vspace{0.1cm}
 & \multicolumn{2}{c}{Planck}&\multicolumn{2}{c}{CMBPol} \\
 & $\mathcal{B}_{i,LD}$ & $r_i$ & $\mathcal{B}_{i,LD}$ & $r_i$\\

\hline
\vspace{0.1cm}
LD       			   & $0.0\pm 0.3$ &  $0.102\pm 0.017$   
					   &$ 0.0\pm 0.3$ & $0.100\pm 0.002$\\
\vspace{0.1cm}
$\quad n_{\rm s}$       & $-6.3\pm 0.3$ & $0.082\pm 0.014$   
						& $-13.0\pm 0.3$ &$0.105\pm 0.001$ \\
\vspace{0.1cm}
$\quad n_{\rm run}$      & $-6.5\pm 0.3$ & $0.086\pm 0.015$
						 & $-15.5\pm 0.3$ &$0.103\pm 0.001$ \\
\vspace{0.1cm}
$\quad 2k_i$               & $-3.1\pm 0.3$  & $0.091\pm 0.015$
				         & $-10.2\pm 0.3$ & $0.101\pm 0.001$\\
\hline
\hline
\end{tabular}
\label{tab:ms}
\end{center}
\end{table}

\section{Discussion and Conclusions}
\label{sec:results}

\begin{figure}
\begin{center}
  \begin{minipage}[c]{0.45\textwidth}
$\begin{array}{cc}
\includegraphics[trim = 30mm 90mm 30mm 80mm, clip, width=7cm, height=5.5cm]{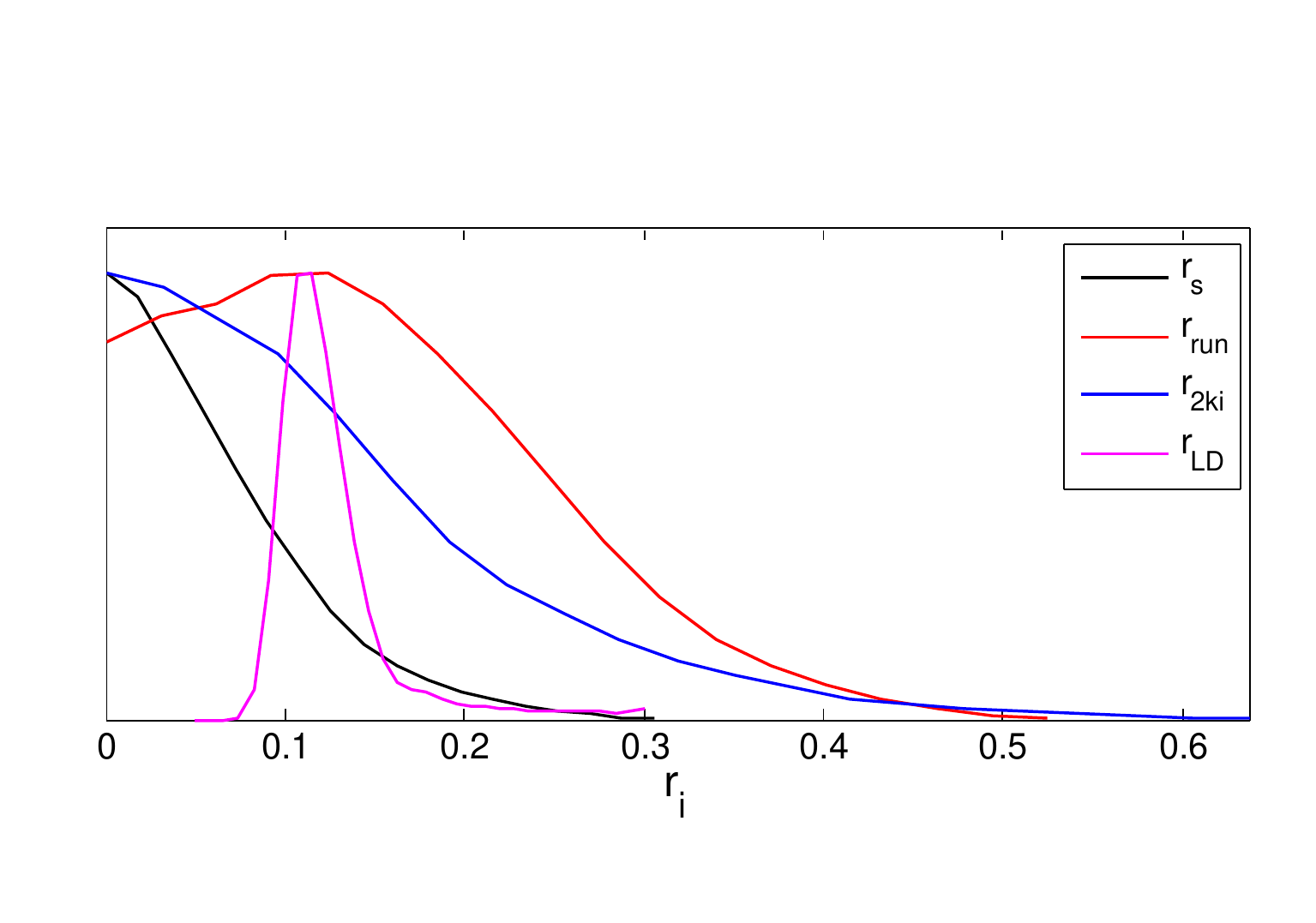}
\end{array}$
  \end{minipage} \qquad
  \begin{minipage}[c]{0.45\textwidth}
  \scalebox{1.0}{
\begin{tabular}{ccc}
\cline{1-3}\noalign{\smallskip}
\hline
\vspace{0.0cm}
\quad Current  & \quad ${\rm N}_{\rm par}$  &$\mathcal{B}_{i,j}$
\\ \hline
\vspace{0.1cm}
$n_{\rm s}$ &               $+3$        &$+0.0\pm 0.3$  \\
\vspace{0.1cm}
$n_{\rm run}$ &     $+4$            &$+2.0\pm 0.3$  \\
\vspace{0.1cm}
$2k_i$ &                $+7$            &$+2.3\pm 0.3$  \\
\vspace{0.1cm}
LD &        $+3$     &$+3.4\pm 0.3$\\
\hline
\hline
\end{tabular}
}
  \end{minipage}

\end{center}
\caption{1-D marginalised posterior distributions of the tensor-to-scalar ratio for
the different models (left panel),  along with their Bayesian evidence and number
of parameters of each model (right panel). The Bayes factor is compared to the 
simple-tilt model ($n_{\rm s}$.) }
\label{fig:all}
\end{figure}

In this paper we have performed a MCMC exploration of the full cosmological 
parameter-space and  showed current and future constraints on the  inflationary 
parameters, with particular attention to the tensor-to-scalar ratio. 
We have considered models that deviate from the standard power-law in the scalar power-spectrum:
a power-law parameterisation with running behaviour and the spectrum predicted from the 
Lasenby \& Doran model. By implementing a model-independent reconstruction for 
$\mathcal{P_R}(k)$, we found that a turn-over in the scalar spectrum
is preferred to explaining cosmological observations.
A similar form of the scalar spectrum has been previously obtained
 assuming different model-independent reconstructions, some of them
 with different data sets \cite{ Guo11, Guo11b, Vazquez12}.
Even though we have not given the results for the standard cosmological
 parameters $\Omega_{\rm b}h^2$, $\Omega_{\rm c}h^2$, $\theta$, $\tau$, their best-fit values 
remained essentially unaffected throughout the models. 
For all the models, we have computed the Bayes factor and compared each to the simple 
power-law parameterisation. We found, using current observations, that the preferred 
model is given by the LD model.
 The summary of the analysis, illustrated in Figure \ref{fig:all},
displays how the constraints on the tensor-to-scalar ratio are broadened for
non-power law models. We observe that the best-fit value of $r_{\rm run}$ is slighly
offset from zero and coincides with the peak of $r_{\rm LD}$. 
It has to be born in mind that if future surveys confirm
small values of the true tensor-to-scalar ratio ($r\lesssim 0.09$), the LD model with a 
$\phi^2$-type potential might be in conflict to reproduce this key feature.
Throughout the models, the tensor-to-scalar ratio has been computed at a particular 
scale $k_0=0.015$ Mpc$^{-1}$. However, to illustrate the robustness of the model
selection, over a different choice of scale $k_0$, we compute the Bayesian evidence 
for all models at $k_0=0.002$ Mpc$^{-1}$. The results are esentially
unaffected and still show a preference for the  LD model. The Bayesian evidece, 
for each model, compared to the power-law parameterisation ($n_{\rm s}$) are as follow:
\begin{eqnarray*}
&\mathcal{B}_{\rm run, n_s}& \qquad \qquad \quad \mathcal{B}_{\rm 2ki, n_s}
      \quad \qquad  \qquad \quad \,\, \mathcal{B}_{\rm LD, n_s}  \\
&+1.8\pm0.3& \qquad \qquad +2.29 \pm 0.3 \qquad \qquad +3.0 \pm 0.3
\end{eqnarray*}

\noindent
We also notice that the tensor-spectrum in the LD model exhibits
a running-like behaviour, contrary to the rest of the models. 
Nevertheless, the addition of a {\it running of the tilt} in the tensor modes,
 $n^t_{\rm run} \equiv d\ln n_{t}/d\ln k$, 
provides no significant changes to the Bayesian evidence.
 This can be seen from the fact that $n^t_{\rm run}$ is nearly zero, 
 and also that no extra-parameters need to be included
due to the existence of a second consistency relation: 
$n^t_{\rm run} \simeq n_{t}[n_t-(n_{\rm s} -1)]$~\cite{Cortes06}. 
For instance, the Bayes factor 
of the $n_{\rm run}$-model with running in the tensor-spectrum is 
$\mathcal{B}_{n_{\rm run}+n^t_{\rm run},\, n_{\rm run}}=+0.4 \pm 0.3 $,
 with constraints $n^t_{\rm run}\times 1000=0.37\pm 0.82$.
A power-law parameterisation of $\mathcal{P_T}(k)$ is therefore 
sufficient to describe current data. We will explore further possibilities in a more
detailed future work.

With regards to future surveys, the Planck satellite will be able to differentiate  
the running and tilt model  from  the LD model, but not  decisively from the 
two-internal-node  reconstruction. The improvement using CMBPol  selects the right 
form of the primordial  spectrum, as shown  in Table   \ref{tab:ms}.

\acknowledgments
This work was carried out largely on the Cambridge High Performance Computing cluster,
DARWIN. JAV is supported by CONACYT M\'exico.

\bibliographystyle{abbrv}
\bibliography{Tensor}




\end{document}